\documentclass[12pt]{article}

\usepackage{times}

\usepackage[round]{natbib}
\usepackage{hyperref}
\hypersetup{colorlinks,
citecolor=blue,
linkcolor=blue,
urlcolor=black
}

\usepackage{algorithm}

\usepackage{mathtools}
\usepackage{setspace}
\usepackage{relsize}
\usepackage{setspace}
\usepackage[margin=1in]{geometry}
\usepackage{fullpage}
\usepackage{multirow}
\usepackage{amsmath, amsfonts, amssymb, amsthm, bm}
\usepackage{bbm}
\usepackage{tikz}
\usepackage{algorithm, algorithmic}
\usepackage{mdwlist}
\usepackage{enumerate}
\usepackage{subcaption}
\usepackage{float}
\usepackage{courier}

\usepackage{makecell}

\makeatletter
\renewcommand*\env@matrix[1][\arraystretch]{%
  \edef\arraystretch{#1}%
  \hskip -\arraycolsep
  \let\@ifnextchar\new@ifnextchar
  \array{*\c@MaxMatrixCols c}}
\makeatother

\newtheorem{theorem}{Theorem}[section]

\newtheorem{lemma}[theorem]{Lemma}
\newtheorem{remark}{Remark}

\newcommand{\real}{\mathbb{R}}

\newcommand{\grad}{{\nabla}}

\allowdisplaybreaks
\makeatletter
\newcommand*{\rom}[1]{\expandafter\@slowromancap\romannumeral #1@}
\makeatother

\newenvironment{proof-of-lemma}[1]{\noindent{\bf Proof of Lemma #1}\hspace*{1em}}{\qed\bigskip}
\newenvironment{proof-of-theorem}[1]{\noindent{\bf Proof of Theorem #1}\hspace*{1em}}{\qed\bigskip}

\begin{document}

\begin{center}
\LARGE An MCMC-free approach to post-selective inference
\end{center}

\newcommand\blfootnote[1]{%
  \begingroup
  \renewcommand\thefootnote{}\footnote{#1}%
  \addtocounter{footnote}{-1}%
  \endgroup
}

\medskip

\begin{center}
{\large Snigdha Panigrahi$^{*}$ \ \ \ Jelena Markovic$^{*}$ \ \ \ Jonathan Taylor }
\end{center}

\begin{center}
	Stanford University
\end{center}

\blfootnote{*To whom correspondence should be addressed (E- mail: snigdha@stanford.edu, jelenam@stanford.edu).}
\begin{abstract}
We develop a Monte Carlo-free approach to inference post output from randomized algorithms with a convex loss and a convex penalty. The pivotal statistic based on a truncated law, called the \textit{selective pivot}, usually lacks closed form expressions. Inference in these settings relies upon standard Monte Carlo sampling techniques at a reference parameter followed by an exponential tilting at the reference. Tilting can however be unstable for parameters that are far off from the reference parameter. We offer in this paper an alternative approach to construction of intervals and point estimates by proposing an approximation to the intractable selective pivot. Such an approximation solves a convex optimization problem in $\mathbb{R}^{|E|}$, where $|E|$ is the size of the active set observed from selection. We empirically show that the confidence intervals obtained by inverting the approximate pivot have valid coverage.
\end{abstract}

\section{Introduction} \label{introduction}

The aim of a selective inference problem is to provide confidence intervals with valid coverage when the same data was used to select the inferential questions of interest. The approach developed in \cite{exact_lasso, optimal_inference} is based on truncating the generative law of the data to realizations that lead to a selection event. The confidence intervals are then obtained by inverting a pivotal statistic based on the truncated law. Previous papers \cite{lee2014exact, exact_lasso, TTLT2015} compute polyhedral constraints on data $y$ that define affine selection rules. They calculate the pivot by applying the CDF transform of a univariate truncated Gaussian law to the target statistic in the saturated model on data $Y \in \real^n$; when $Y\lvert X \sim \mathcal{N}(\mu, \Sigma)$ for a fixed $X\in \real^{n \times p}$. The pivot is however intractable while attempting to provide inference after a randomized selection as in \cite{randomized_response} and even, in non-randomized settings for more general generative models like the selected model in \cite{optimal_inference}. Thus, the problem of inverting the pivot to obtain confidence intervals is much harder in more general models and randomized settings. Our methods in the current paper offer an approximation to the intractable pivot as a function of the parameters in the generative model. This allows us to invert the approximate pivot directly to obtain confidence intervals as opposed to an MCMC sampling from the truncated law at a reference parameter. 

We develop tools to provide valid inference in the truncated framework after observing an active set $E$ with signs $s_E$ from solving randomized algorithms with a convex loss and a convex penalty. Randomization is introduced as a linear term in the objective of a constrained learning program as in \cite{harris2016selective}. This leads to a selection on a perturbed version of the data but, preserves more left-over information for inference; see \cite{randomized_response}. We refer to a pivotal statistic based on the truncated law that is inverted to obtain confidence intervals as the \textit{selective pivot}. Deferring details to the technical sections of the paper, the selective pivot based on 
	a test statistic that unconditionally satisfies $T^{obs}\sim\mathcal{N}(b,\sigma^2)$ takes the form
\begin{equation*}
p(T^{\text{obs}}; b, \sigma) = \cfrac{\int_{T^{\text{obs}}} ^{\infty} \exp(-(t- b)^2/2 \sigma^2) h(t) dt}{\int_{-\infty} ^{\infty} \exp(-(t- b)^2/2 \sigma^2) h(t) dt}.
\end{equation*}
 The function $h(\cdot)$ is the volume of an affine region with respect to a multivariate law and thus, lacks a closed form expression. Our key contribution is the proposal of a convex approximation $\hat{h}(\cdot)$ to $h(\cdot)$ and using to compute $p(T^{\text{obs}}; b, \sigma)$ on a grid $G$ in the real line. 
We invert the approximate selective pivot to construct confidence intervals in Section \ref{experiments}, which are empirically seen to have the target coverage. This validates the accuracy of our approximation. Our methods also enjoy the higher statistical power inherited from randomization in the selection stage. Confidence intervals based on the construction in \cite{exact_lasso} are known to grow very wide when the observed statistic is close to the selection boundary. Whereas, our intervals have comparable lengths to the unadjusted intervals due to randomized selection.

The idea behind the approximation is smoothening an upper bound to an intractable multivariate probability. This results in solving a convex optimization problem in $\mathbb{R}^{|E|}$, where $|E|$ is the size of the active set chosen by the randomized program. An MCMC approach on the other hand is based on sampling from a reference distribution, but this is not enough to obtain confidence intervals. One has to employ exponential tilting at the reference parameter to obtain confidence intervals; this can lead to unstable pivots at parameters far off from the reference. Our approach provides a direct computation of pivots without implementing any sampler making it free from sampler error. An additional advantage of our method is that we can maximize the approximate truncated law to compute the selective MLE. 
This is possible as we approximate the normalizer on a grid $G\subset \real$ as
\begin{equation*}
\int_{-\infty} ^{\infty} \exp(-(t- b)^2/2 \sigma^2) h(t) dt\approx \sum_{t \in G} \exp(-(t- b)^2/2 \sigma^2) \hat{h}(t).
\end{equation*}
 The MLE from the approximate truncated law can also be used as a reference for MCMC samplers targeting conditional inference. We note that our methods can be parallelized while computing intervals for $|E|$ variables and also, while computing $\hat{h}$ along a grid. This shall become clear after the details of the algorithm. 
Such a parallel computation is harder for samplers that need the previous draw to implement a new draw.

Our approach can be generally viewed as a pseudo-likelihood approach used in different problem settings earlier in \cite{wolfinger1993generalized, liang2003maximum, chen2005pseudo}. Related works in the selective inference literature are \cite{bootstrap_mv}, which constructs confidence intervals via Monte Carlo sampling in the randomized settings. In the non-randomized realm of selective inference, \cite{yang2016selective} construct one-sided and conservative confidence intervals for more general parameters after group-sparse selection methods. Revisiting proposals on point estimation post selection, \cite{reid_many_means} computes the MLE based on a univariate truncated likelihood, a reduction possible again in the simpler sequence models. \cite{selective_bayesian} uses a similar technique of smoothening a Chernoff bound to approximate affine Gaussian probabilities. 

The rest of the paper is organized as follows. Section \ref{background} presents the truncated law after solving a randomized convex program, reviewing some of the previous work in \cite{harris2016selective}. It introduces the selective pivot that is inverted to construct confidence intervals and presents a motivating example based on the methods in the paper. Section \ref{approximate:pivot} states the main technical results of the paper that lead to computation of an approximate selective pivot based on the truncated law. It outlines the algorithm employed for solving an optimization problem linked with the approximation; optimization solves a convex objective in $|E|$ dimensions for each population coefficient in the generative model. Section \ref{experiments} applies our approach to both simulated data and real data to construct valid confidence intervals post some popular selection procedures and compares the adjusted estimates against those from the untruncated law.\footnote{By untruncated law we mean naive intervals based on Gaussian quantiles from the unconditional distribution that ignores selection.}

\section{Technical background and motivation}
\label{background}

\subsection{Targets and generative models}

We provide inference for an adaptive target chosen after solving a randomized convex program based on data $(Y\in \real^n , X\in \real^{n \times p})$ as
\begin{equation} \label{canonical:randomized:program}
\begin{aligned}
{\hat{\beta}(X,y,\omega) = \underset{\beta\in\mathbb{R}^p}{\textnormal{argmin}}\: \ell(\beta;(X,y)) + \mathcal{P}(\beta) -\omega^T \beta} + \frac{\epsilon}{2}\|\beta\|_2^2,
\end{aligned}
\end{equation}
where $(X,y)\times \omega\sim F\times G$. The linear term in $\omega$ inducts randomization into the objective of the problem, where $\omega\sim G$ with a density $g$ supported on $\mathbb{R}^p$. For some algorithms like the Lasso that do not always guarantee a solution, a typical $\ell_2$ penalty term for small $\epsilon>0$ as in \cite{elastic_net} is included in the objective. The fact that the same data that was used to select the target of interest is now being used for inference invalidates intervals based on the untruncated model. We review the main concepts of the selective inferential framework through an example of a logistic lasso problem.

Consider selecting a model using in \eqref{canonical:randomized:program} a logistic loss function 
\begin{equation*}
\ell(\beta;(X,y))=-\sum_{i=1}^n \left(y_i\log\pi(x_i^T\beta)+(1-y_i)\log(1-\pi(x_i^T\beta))\right),
\end{equation*}
where $\pi(x)=e^x/(1+e^x)$; $x_i$, $i=1,\ldots, n$, are the rows of $X$ and $\mathcal{P}(\beta)=\lambda\|\beta\|_1$ is an $\ell_1$-penalty function. %
Denote the set of selected variables as $E$ with their corresponding signs given by $s_E$ and the minimizer of \eqref{canonical:randomized:program} with logistic loss as $\hat{\beta} = (\hat{\beta}_E,0)$. The selection event observed from the output of the above program is 
	\[\{(X,y,\omega):\hat{\beta}_{-E}=0 \;\textnormal{ and }\; \textnormal{sign}(\hat{\beta}_E) = s_E\}\]
which is equivalent to observing $(E, s_E)$ as solutions of solver in \eqref{canonical:randomized:program}. We condition additionally on the signs of the active variables to get a polytope as the selection region as in \cite{lee2014exact, exact_lasso}, else we get a union of polytopes.
An adaptive target for inference based on knowing $E$ is the population coefficients $b_{E}$ that satisfies 
\begin{equation*}
\mathbb{E}_F[X_{E}^T\left(y-\pi_{E}(b_{E})\right)]=0,
\end{equation*}
where $\pi_{E}(b_{E})= {\exp(X_{E}b_{E})}/{(1+\exp(X_{E}b_{E}))}$.
If $T_E = \bar{\beta}_E$ is the MLE solution to the untruncated logistic law involving only predictors in $E$ and $T_{j\cdot E}$ is the $j$th coordinate of $T_E$, then unconditional inference on $b_{j\cdot E}$, $j\in E$  is based on the asymptotic distribution of $T_{j\cdot E}=\bar{\beta}_{j\cdot E}$. Standard asymptotics tell us that $T_{j\cdot E}$, called the target statistic, has an unconditional asymptotic law 
\begin{equation*}
	T_{j\cdot E}-b_{j\cdot E}\stackrel{d}{\rightarrow} \mathcal{N}(0,\sigma_j^2)
\end{equation*}
 as  $n\rightarrow\infty$, where $\sigma_j$ denotes the asymptotic variance. 
But, since the choice of $b_E$ is made only after observing the active set $E$ from \eqref{canonical:randomized:program}, inference using usual asymptotic Gaussian law is no longer valid.

To validate inference, we consider the distribution of $T_{j\cdot E}$ conditional on observing $(E, s_E)$. Inference about $b_E$ requires us to assume a model on our data $(X,y)$. We choose to work with a \textit{saturated model framework} in the current work, this means that we impose no additional restrictions on the data generating distribution $F$. We point out that our methods are flexible enough to extend to other targets and other generative models which can be guided by selection.
\begin{remark}
\emph{\textit{\textbf{Other generative models and targets:}}}
Another commonly used generative family of models is $\{F : \mathbb{E}_F[Y \lvert X] = X_E b_E\},$
called the ``selected model'' for inference where we impose conditions on the conditional mean of $Y\lvert X$ upon selection \citep{optimal_inference}. Our framework of methods offers the flexibility to extend inference to others parametrization of the conditional mean, example we may assume $\mathbb{E}_F[Y \lvert X]  = X_{\bar{E}}b_{\bar{E}}$ as a plausible generating model, where $\bar{E}$ is determined only through $E$. 
Upon seeing the selected model $E$, an analyst based on her expertise, can decide to report the coefficients corresponding to $\tilde{E}$ that may not necessarily agree with $E$. In that case, she would report the confidence intervals for $b_{\tilde{E}}$ satisfying $\mathbb{E}_F[X_{\tilde{E}}^T(y-\pi_E(b_{\tilde{E}}))]=0$. 
\end{remark}

\subsection{Selective pivot based on a change of measure}

Having described the adaptive target of interest and the model on data, we turn attention to the truncated law of the data conditional on selection. This is the generative model on the data truncated to realizations leading to the same selection event.
The subgradient equation of \eqref{canonical:randomized:program} with an $\ell_1$ penalty term $\mathcal{P}(\beta) =\lambda \|\beta\|_1$ gives a change of measure formula 
\[\omega = \nabla\ell(\hat{\beta};(X, y))+  \begin{pmatrix} \lambda s_E \\ u_{-E}\end{pmatrix} +\epsilon\begin{pmatrix}
		\hat{\beta}_E \\ 0 \end{pmatrix},\]
where $\hat{\beta}$ is the solution to \eqref{canonical:randomized:program} and $u_{-E}$ is the subgradient vector the penalty corresponding to inactive variables (the ones not in $E$). Denote as $O = \begin{pmatrix}
	\hat\beta_E \\ u_{-E} \end{pmatrix}$ and call $O$ the optimization variables. 
Denote the observed data vector $D= \begin{pmatrix} D_E\\ D_{-E}\end{pmatrix} = \begin{pmatrix} \bar{\beta}_E \\ X_{-E}^T(y-\pi_E(\bar{\beta}_E)) \end{pmatrix}$, where $\bar{\beta}_E$ is the MLE of the untruncated logistic problem involving only predictors in $E$. 

A Taylor expansion of the gradient of the loss yields a linear map between randomization and the augmented vector $(D,O)$, called \textit{randomization reconstruction} as 
\[\text{(\textbf{RR})}:\; \omega = \omega(D,O)= A_0D + BO+\gamma.\]
where $A_0$ and $B$ are fixed matrices and $\gamma$ is a fixed vector. 
The detailed derivations with explicit expressions for $A_0, B$ and $\gamma$ are in the supplement.
The selection of $(E, s_E)$ from the solver in \eqref{canonical:randomized:program} is described by the map $\omega(D,O)$ where 
optimization variables $O$ are constrained to the region \[\mathcal{K}=\{o\in\mathbb{R}^p:\textnormal{sign}(o_E)=s_E, \|o_{-E}\|_{\infty}\leq\lambda\}.\]

Selective inference is based on the joint law of data and randomization $(D,\omega)$, conditional on the event that constrains the optimization variables $O$ to lie in $\mathcal{K}$. A change of measure formula \citep{harris2016selective} from the space of $(D, \omega)$ to that of $(D, O)$ enables to sample from a density supported on the much simpler constraint region $\mathcal{K}$, tensors of orthants and cubes as in the Lasso problem outlined here. Using the change of measure trick of \cite{harris2016selective}, the truncated joint density of $(D,O)$ at $(d,o)$ becomes 
\[f_D(d)\cdot g(\omega(d,o))\cdot \mathbb{I}_{\{o\in\mathcal{K}\}},\]
where $f_D(D)$ is the pre-selection density of $D$, an asymptotic Gaussian.
To provide inference for $\beta_{j\cdot E}$, recall that the target statistic is $T_{j\cdot E}=\bar{\beta}_{j\cdot E}$. We decompose the affine map in data vector $D$ given by $A_0 D$ in $\omega(D,O)$ into a part involving the target statistic and a component involving nuisance parameters. Using the joint asymptotic normality of $T_{j\cdot E}$ and $D$ for inference, we do data decomposition of $A_0 D$ into asymptotically independent components as \[\text{(\textbf{DD})}:  \;A_0 D =  A_0 \Sigma_{D, T_{j\cdot E}}T_{j\cdot E}/\sigma_j^2 +  A_0 (D-\Sigma_{D,T_{j\cdot E}}T_{j\cdot E}/\sigma_j^2) = A_jT_{j\cdot E} + F_j,\]
where $\Sigma_{D,T_{j\cdot E}}$ denotes the asymptotic cross-covariance of data vector and target statistic, 
\[A_j=A_0 \Sigma_{D, T_{j\cdot E}}/\sigma_j^2 \;\;\text{ and }\;\; F_j=A_0(D-\Sigma_{D,T_{j\cdot E}}T_{j\cdot E}/\sigma_j^2).\] We condition additionally on $F_j$ in the conditional law since the asymptotic distribution of $F_j$ involves nuisance parameters; see \cite{optimal_inference} for more details. 
$\Sigma_{D,T_{j\cdot E}}$ and $\sigma_j^2$, used in the above decomposition, are easily estimable using pairs bootstrap in most cases; see \cite{bootstrap_mv} for more details on the decomposition map for inference on general targets.

With the above decomposition, the (asymptotic) truncated density of $(T_{j\cdot E}, O)$ given the nuisance parameters $F_j$ in a saturated model at a realization $(t_{j\cdot E},o)$ is proportional to 
\[
\exp(-\|t_{j\cdot E} - b_{j\cdot E}\|^2_2/2\sigma_j^2)\cdot g(A_jt_{j\cdot E} + B o +  c_j) \cdot \mathbb{I}_{\{o \in \mathcal{K}\}}, 
\]
 where $c_j=F_j^{\text{obs}} + \gamma$ with $F_j^{\text{obs}}$ as the observed value of statistic $F_j$.
The marginal density of the target statistic conditional on the selection event and nuisance statistic, marginalizing over $O$, denoted as $f(t_{j\cdot E}\lvert \hat{E} = E, F_{j\cdot E})$ is proportional to 
\begin{equation} \label{selective:marginal}
\exp(-(t_{j\cdot E} - b_{j\cdot E})^2/2\sigma_j^2)\cdot h(t_{j\cdot E}),
\end{equation}
where
\[
h(t_{j\cdot E})=\mathbb{P}(O\in \mathcal{K} \lvert T_{j\cdot E}= t_{j\cdot E}) =\int\limits_{o\in \mathcal{K}}g(A_jt_{j\cdot E} + B o +  c_j)do.
\] 
We refer to this density as \textit{selective marginal} density of $T_{j\cdot E}$.
Thus, the target selective law decouples into the pre-selection density of the target statistic and the selection probability of $O \in \mathcal{K}$ given $T_{j\cdot E}$, that is free of parameter $b_{j\cdot E}$. Note that, $\mathbb{P}(O\in \mathcal{K} \lvert T_{j\cdot E}= t_{j\cdot E})$ is the volume of an affine region $\mathcal{K}$ with respect to a multivariate law in $\real^p$. Hence, it does not have an easily available closed form expression. The implication is that if we know the function $h(t) = \mathbb{P}(O\in \mathcal{K} \lvert T_{j\cdot E}= t)$, we have to compute it only once to obtain the selective marginal density at any parameter value $b_{j\cdot E}$.

To summarize the discussion above, the generative model conditional on selection and appropriate statistics to eliminate nuisance parameters forms the truncated law. The \textit{selective pivot} for the parameter $b_{j\cdot E}, j\in E$, using the selective marginal density of the target statistic $T_{j\cdot E}$ in \eqref{selective:marginal} is
\[p(T_{j\cdot E}; b_{j\cdot E}, \sigma_j)=\cfrac{\int_{T_{j\cdot E}} ^{\infty} \exp(-(t- b_{j\cdot E})^2/2 \sigma_j^2) h(t) dt}{\int_{-\infty} ^{\infty} \exp(-(t- b_{j\cdot E})^2/2 \sigma_j^2) h(t) dt}.\]
Selective pivots based on the non-randomized version of this problem (without the randomization term in the objective) have been considered in \cite{penalized_l1} and the randomized analog in \cite{randomized_response}. 
What we are aiming for in this work is an approximation to $h(t) =  \mathbb{P}(O\in \mathcal{K} \lvert T_{j\cdot E}= t).$
 A calculation of the approximate $h(\cdot)$ on a grid in the real line leads to an \textit{approximate selective pivot}.

\subsection{A motivating example}

Before we discuss our method of approximating the intractable function $h(t) =  \mathbb{P}(O\in \mathcal{K} \lvert T_{j.\cdot E}= t),$ we present below an example where selection is performed using a randomized program with the logistic loss and an $\ell_1$ penalty. The experiment is described as follows. The data $(X \in \real^{n \times p} ,\; y \in \real^n)$, with $n=1000$ $p=500$,
is generated as $x_i \stackrel{ind}{\sim} \mathcal{N}(0,I_p), y_i\stackrel{ind}{\sim} \mathcal{N}(0,1)$ with columns of $X$ normalized. After observing $(E,s_E)$, the target parameter of interest is $b_E$ and the target statistic $\bar{\beta}_E$.
To provide inference for the target $b_E$ post the output of solver in \eqref{canonical:randomized:program} in the above \textit{all noise} model, we compare the naive and selective pivots and intervals (see Figure \ref{fig:LG}). Naive (untruncated) approach bases inference for $b_{j\cdot E}$, the $j$th coordinate of the population coefficient based on the naive Gaussian pivot $1-\Phi((T_{j\cdot E}- b_{j\cdot E})/\sigma_j)$. Approximate pivot for $b_{j\cdot E}$ is based on the approximate selective pivot 
\begin{equation*}
\hat{p}(T_{j\cdot E};b_{j\cdot E}, \sigma_j)=\cfrac{\sum_{t\in G:\:  t\geq T_{j\cdot E}}\exp(-(t_{j\cdot E} - b_{j\cdot E})^2/ 2\sigma_j^2)\cdot \hat{h}(t) }{\sum_{t\in G}  \exp(-(t_{j\cdot E} - b_{j\cdot E})^2/2\sigma_j^2)\cdot \hat{h}(t)}
\end{equation*}
 using our method of approximating $h(\cdot)$ as $\hat{h}(\cdot)$ on a grid $G$ in the real line. 

\begin{figure}[H]%
    \centering
    	\includegraphics[height=8cm, width=9cm]{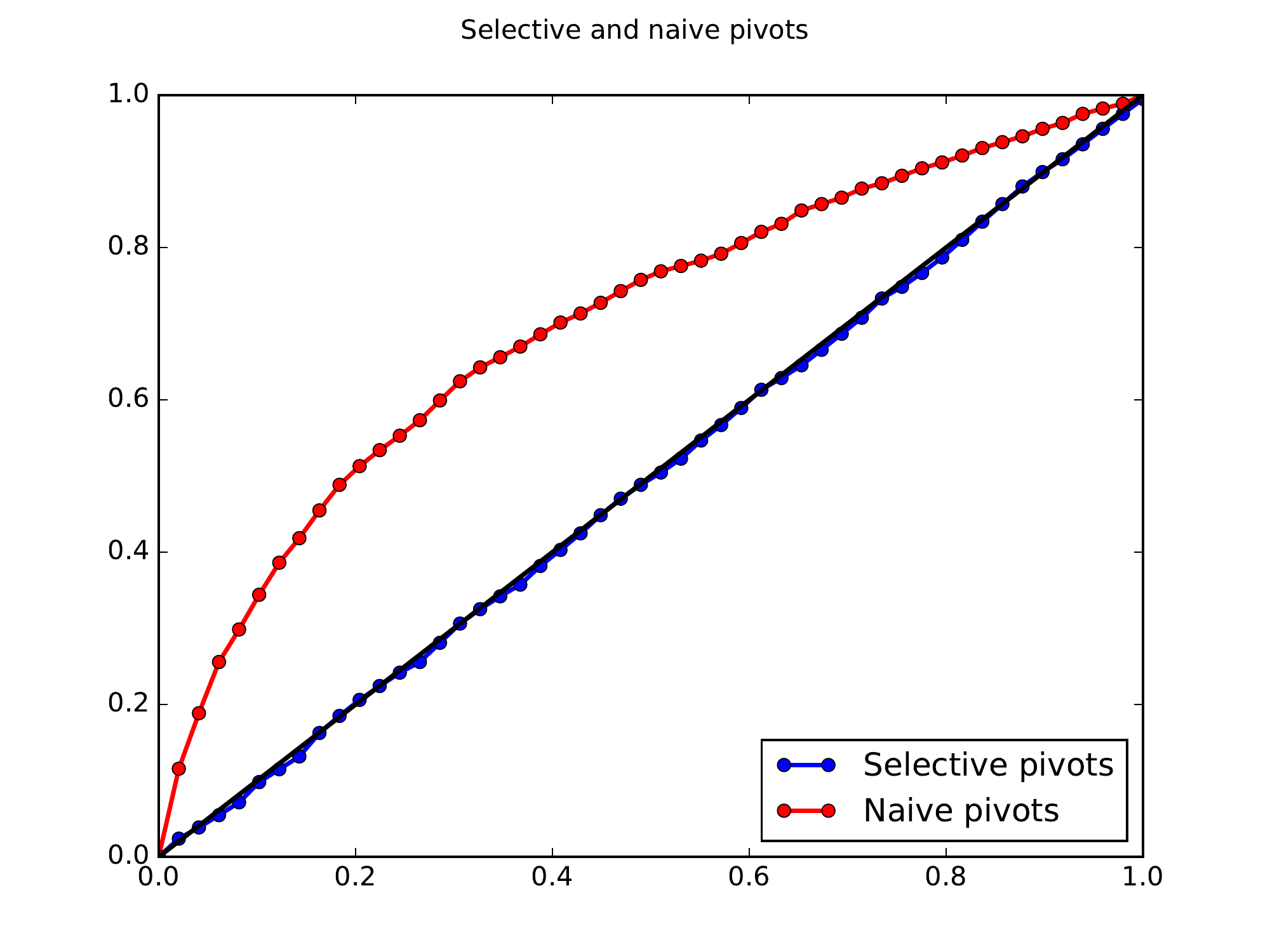}
    \caption{The plot compares the approximate pivots (blue) targeting the true conditional law against the naive pivots (red) as described above based on 100 replications of the experiment. The approximate  selective pivot is uniform as opposed to the naive one. Furthermore, the intervals based on our approximate pivots modeled on the conditional approach cover the true target $b_E$ $90\%$ of the times as opposed to a mere coverage of $66\%$ of the naive Gaussian intervals based on the naive approach.}
    \label{fig:LG}
\end{figure}


\section{Approximate selective pivot} \label{approximate:pivot}
\subsection{Approximate affine volume}

In this section, we propose an approximation to volumes of affine regions with respect to a multivariate law. The intractable probability that we are after is $h(t) =\mathbb{P}(O\in \mathcal{K} \lvert T_{j\cdot E}= t) $
in \eqref{selective:marginal} to construct the selective pivot based on target statistic $T_{j\cdot E}$. This provides a sampling-free alternative to compute $p$-values, that can be inverted directly to obtain confidence intervals and point estimates like the selective MLE.
We present results in this section for a Gaussian randomization, that is $\omega \sim \mathcal{N}(0, \tau^2 I_p)$ in \eqref{canonical:randomized:program} with $p$ independent components. These results also extend easily to other multivariate randomizations densities like the Laplace and Logistic etc. The next section on examples shows that the structure of the logistic lasso problem carries through for other interesting selection problems with different losses and different penalties and two randomization schemes: Gaussian and Laplace.

Our technique of approximating $h(\cdot)$ is involves smoothening an upper bound through a barrier penalty, which yields a smooth approximation to $h(\cdot)$. Our approach is similar to \cite{selective_bayesian} which uses a smooth version of a Chernoff bound; the upper bound used in the current paper is however different from the usual Chernoff bound. The approximation is stated in the next theorem and lemma, the key technical results of this paper that lead to an approximate selective pivot . Before stating the following theorem, we further partition reconstruction map  into active and inactive parts
\[\text{(\textbf{PR})}: \;A_j=\begin{pmatrix}
	A_{j, E} \\ A_{j, -E}
\end{pmatrix}, \;\; B=\begin{pmatrix}
	B_E\\ B_{-E}
\end{pmatrix}, \;\; c_j=\begin{pmatrix}
	c_{j,E}\\ c_{j,-E}\end{pmatrix}\]
where $A_{j, E}$ is the submatrix containing $E$ rows of $A_j$ and similar decomposition goes for $B$ and $c$. 

\begin{theorem} \label{thm:upper:bound}
\textit{\textbf{An upper bound:}} With an isotropic centered Gaussian randomization with variance $\tau^2 I_p$, an upper bound for $\log\mathbb{P}(O\in \mathcal{K} \lvert T_{j\cdot E}=t_{j\cdot E})$, the volume of selective region $\mathcal{K}$ as stated above, computed with respect to the density of $O$ can be computed as
\begin{equation} \label{chernoff:upper:bound}
\begin{aligned}
&-\sup\limits_{\beta}\biggl\{\inf\limits_{\textnormal{diag}(s_E){u}\geq 0}\Big\{\beta^T u -H(u; t_{j\cdot E})\Big\}- \log \mathbb{E}\left[\exp(\beta^T O_E)\:|\:T_{j\cdot E}=t_{j\cdot E}\right]\biggr\}, 
\end{aligned}
\end{equation}
with
\[ \text{(\textbf{CP})}: \;H(u; t_{j\cdot E})=\sum\limits_{i=1}^{p-|E|} \log \left(\Phi\left((\lambda+ \alpha(u; t_{j\cdot E})_i)/\tau\right) - \Phi\left((-\lambda+ \alpha(u; t_{j\cdot E})_i)/\tau\right)\right)\]
where $\alpha:\mathbb{R}^{|E|}\rightarrow\mathbb{R}^{p-|E|}$ defined as $\alpha(u; t_{j\cdot E}) =  A_{j,-E}t_{j\cdot E}  + B_{-E} u + c_{j,-E}.$
\end{theorem}

Using minimax theorem, we further approximate the upper bound above to get the following \textit{approximation:} for $\log{\mathbb{P}}(O\in\mathcal{K}|T_{j\cdot E}=t_{j\cdot E})$ as
\[\sup\limits_{{\text{diag}(s_E) o_E>0}} -\{\| A_{j,E}t_{j\cdot E} + B_E o_E + c_{j,E} \|_2^2/2\tau^2 - H(o_E; t_{j\cdot E})\}.\]
The details of the proof of Theorem \ref{thm:upper:bound} and the approximation above are in the supplement. We make some crucial observations about the above approximation related to the cube probability (\textbf{CP}), $H(\cdot; t_{j\cdot E})$, and the dimension of the optimization problem. Note that the function $H(\cdot; t_{j\cdot E})$ is the logarithm of the probability of a Gaussian random variable  $\mathcal{N}(-\alpha(o_E,; t_{j\cdot E}), I_{p-|E|})$ lying inside a cube $[-\lambda, \lambda]^{p-|E|}.$ This appears from integrating the inactive sub-gradient variables over the cube. We see that the above approximation solves an optimization problem in $E$ dimensions imposing signs constraints of having observed $s_E$. It is free of the regression dimension $p$ and the sample size $n$ since $|E|$ is usually much smaller than $n$ and $p$.


The RHS of the stated approximation is a constrained optimization problem which can be modified with a barrier penalty inside the constraint region. This yields a smoother approximation cast as an unconstrained optimization, called \textit{smooth approximation}. This is defined as the result of the following optimization over active constraints, yielding approximate $ \hat{h}(t_{j\cdot E})$ as
\begin{equation} \label{barrier:approx}
\text{(\textbf{OA})}:\;\exp\left(\sup\limits_{o_E}-\left\{ \|A_{j,E} t_{j\cdot E}+ B_E o_E + c_{j,E} \|_2^2/2\tau^2- H(o_E; t_{j\cdot E}) + \mathcal{B}(o_E)\right\} \right)
\end{equation}
where $\mathcal{B}$ is a choice of barrier function representing the sign constraints on $o_E$, that is $\text{sign}(s_E)o_E>0$.
This latter formulation gives an unconstrained problem with a smooth, continuous penalty replacing a $0-\infty$ version.
Thus, we approximate the selective pivot $p(T_{j\cdot E}; b_{j\cdot E},\sigma_j)$  using \eqref{barrier:approx}. 
\begin{remark}
The barrier function used in our implementations is given by $\mathcal{B}(o_E) = \log\left(1+ (s_E o_E)^{-1}\right)$, coordinate-wise.
\end{remark}

\subsection{Selective inference: point estimates and intervals}

In this section, we describe inference based on the approximate selective pivot using \eqref{barrier:approx}. It can be used to compute the approximate normalizer to the truncated marginal law in \eqref{selective:marginal} as a function of $b_{j\cdot E}$ on a grid in $\mathbb{R}$. This enables us to compute $p$-values, intervals and point estimates like the selective MLE fairly directly. Denoting 
$\hat{h}(t_{j\cdot E}) = \widehat{\mathbb{P}}(O\in \mathcal{K} \lvert T_{j\cdot E}=t_{j\cdot E})$, the pivotal statistic applies the CDF transform of the law $f(t_{j\cdot E}\lvert \hat{E} = E, F_{j\cdot E} )$, approximated using \eqref{barrier:approx} to the observed statistic value $T_{j\cdot E}$. The approximation of $f(t_{j\cdot E}\lvert \hat{E} = E, F_{j\cdot E} )$, called pseudo-likelihood, is given by
\begin{equation*}
\hat f(t_{j\cdot E}\lvert \hat{E} = E, F_{j\cdot E} )=\cfrac{ \exp(-(t_{j\cdot E} - b_{j\cdot E})^2/2\sigma_j^2)\cdot \hat{h}(t_{j\cdot E})}{\sum\limits_{t\in G}\exp(-(t- b_{j\cdot E})^2/2\sigma_j^2)\cdot \hat{h}(t)}.
\end{equation*}
The denominator is the normalizer computed on a grid $G$ of values of $t$ with approximation $\hat{h}$. The selective pivot $p(\cdot; b_{j\cdot E})$ can thus be approximated on a grid as
\[\hat p(t_{j\cdot E}; b_{j\cdot E}, \sigma_j)= \sum_{t\in G:\:t\geq t_{j\cdot E}} \hat{f}(t\lvert \hat{E} = E, F_j)\Big/{\sum_{t \in G} \hat{f}(t\lvert \hat{E} = E, F_j)}.\]
In simulations, we compute a two sided $p$-value as $\hat{P}(t_{j\cdot E}; b_{j\cdot E}, \sigma_j) = 2\cdot \min(\hat p(t_{j\cdot E}; b_{j\cdot E}, , \sigma_j), 1- \hat p(t_{j\cdot E}; b_{j\cdot E}, \sigma_j))$.
The $100(1-\alpha)\%$ two-sided confidence intervals calculated by inverting the approximate selective pivot are also straight-forward, given by $\{b_{j\cdot E}\in \real :  \hat{P}(t_{j\cdot E}; b_{j\cdot E}, \sigma_j)  \leq \alpha\}$.
In the absence of a hand on the normalizer, the selective MLE in the randomized setting is intractable. But, equipped with the approximation, the next lemma gives an estimating equation for the selective MLE of the population coefficients on the same grid in $\mathbb{R}$. The quantity we use is gradient of the approximate negative log-likelihood (\textbf{GL}) denoted as $\nabla\mathcal{L}$. A standard gradient descent algorithm allows an iterative computation of the same. The proof of below lemma is outlined in the supplement.

\begin{lemma} \textit{\textbf{Selective mle:}}
\label{sel:MLE}
The selective MLE, $\hat{b}_{j\cdot E}$, for $b_{j\cdot E}$ based on approximation \eqref{barrier:approx} on a grid $G$ in $\mathbb{R}$ satisfies an estimating equation
\begin{equation*}
	t_{j\cdot E} { \left\{\sum_{t\in G}\exp((-t^2/2 + \hat{b}_{j\cdot E} t)/\sigma_j^2)\cdot \hat{h}(t)\right\}}=   \sum_{t\in G}t\cdot\exp((-t^2/2 +\hat{b}_{j\cdot E} t)/\sigma_j^2)\cdot \hat{h}(t).
\end{equation*}
\end{lemma}


\subsection{Algorithm for inference}
\label{solver}

Algorithm \ref{alg:p:values:conf:int} computes the approximate pivot $\hat{p}(t_{j\cdot E};b_{j\cdot E}, \sigma_j)$ for parameter $b_{j\cdot E},j\in E$. A two sided $ p$-value $\hat{P}(t_{j\cdot E}; b_{j\cdot E}, \sigma_j)$ for ${b_{j\cdot E}}$ and a two-sided $100(1-\alpha)\%$ confidence interval for 
$b_{j\cdot E}$ can be computed directly from $\hat{p}(t_{j\cdot E};b_{j\cdot E}, \sigma_j)$, as described in Section \ref{approximate:pivot}. Algorithm \ref{alg:mle} computes the selective MLE $\hat{b}_{j\cdot E}$, $j\in E$. The step size of the gradient descent is denoted as $\eta$ and a tolerance to declare convergence is denoted as ``tol.''

\begin{algorithm}[H] \caption{Approximate selective pivot for $b_{j\cdot E}$ } \label{alg:p:values:conf:int}
\begin{spacing}{1.5}
\begin{algorithmic} 
\REQUIRE $\ell(\beta;(X,y))$, $\lambda$, $\epsilon>0$, $\tau^2$, $\alpha$, $G$
\ENSURE $S^*\subset E$
\STATE {\bf(RR)}: $\omega= A_0 D + B O + \gamma$
\STATE {\bf(DD)}: $A_0 D =  A_jT_{j\cdot E} +  F_j$
\STATE {\bf(PR)}: 
$\omega =  \begin{pmatrix} A_{j,E} t_{j\cdot E}+ B_Eo_E + c_{j,E} \\   \alpha(o_E; t) + o_{-E}\end{pmatrix}$
\FORALL {$t$ in grid $G$}
\STATE {\bf{(CP)}}:  $H(o_E; t) = \sum\limits_{i=1}^{p-E}\log (\Phi((\lambda+ \alpha(o_E; t)_i)/\tau) -\Phi((-\lambda+ \alpha(o_E; t)_i)/\tau)) $
\STATE{\bf{(OA)}}: $\log \hat{h}(t) = -\underset{o_E}{\inf}\{\|A_{j, E}t+ B_E o_E + c_{j,E} \|_2^2/2\tau^2- H(o_E;t) + \mathcal{B}(o_E)\}$
\ENDFOR
${\;\hat{p}(t_{j\cdot E};b_{j\cdot E}, \sigma_j)=\frac{\sum_{t\in G:\: t\geq t_{j\cdot E}}\exp\left(-(t - b_{j\cdot E})^2/2\sigma_j^2\right)\cdot \hat{h}(t)}{\sum_{t\in G}\exp\left(-(t- b_{j\cdot E})^2/2\sigma_j^2\right)\cdot \hat{h}(t)}}$.
\end{algorithmic}
\end{spacing}
\end{algorithm}


\begin{algorithm}[H] \caption{Selective MLE} \label{alg:mle}
\begin{spacing}{1.5}
\begin{algorithmic} 
\REQUIRE $\ell(\beta;(X,y))$, $\lambda$, $\epsilon$, $\tau^2$, $\eta$, tol.
\ENSURE $S^*\subset E$
\STATE Repeat {\bf(RR)}, {\bf(DD)}, {\bf(PR)}
\FORALL {$t$ in grid $G$}
\STATE {\bf{(CP)}} and {\bf{(OA)}}
\ENDFOR
\WHILE{ $\hat{b}_{j\cdot E}^{(K)} - \hat{b}_{j\cdot E}^{(K-1)}>\text{tol.}$}
\STATE $\hat{b}_{j\cdot E}^{(K)} = \hat{b}_{j\cdot E}^{(K-1)} -\eta \cdot \grad(\mathcal{L}(b_{j\cdot E}^{(K-1)}))$
\STATE {\bf{GL}}: $\grad \mathcal{L}(b_{j\cdot E}^{(K)})$ at $K$th iteration
\ENDWHILE
\end{algorithmic}
\end{spacing}
\end{algorithm}

\section{Experiments} \label{experiments}

As experiments, we construct confidence intervals correcting for selection after running various selection procedures with different losses and penalties and randomization distributions. The below table gives the loss functions and penalties of the selection procedures implemented in this section: forward stepwise (FS), Lasso and logistic Lasso 

\begin{center}
\small
\captionof{table}{Losses and penalties}
\label{table:models}
\bgroup
\def\arraystretch{2.0}
\begin{tabular}{ |c|c|c|c|c| }
\hline
\bf{Algorithm} & \bf{Loss} &  \bf{Penalty} \\
\hline
\bf{FS} & $-\beta^T X^T y$ & $I^{1}_{\ell_{1}}(\beta)$ \\
\hline
\bf{Lasso} & $\frac{1}{2}\|y- X\beta\|_2^2$  & $\lambda \|\beta\|_1$ \\
\hline
\thead{Logistic\\Lasso} & \thead{$-\sum_{i=1}^n (y_i \log\pi(x_i^T\beta)$ \\ $+ (1-y_i)\log(1-\pi(x_i^T\beta))$}  & $\lambda \|\beta\|_1$ \\
\hline
\end{tabular}
\egroup
\end{center}
The penalty in FS is the characteristic function of $\ell_1$ unit ball:
\begin{equation*}
I^1_{\ell_{1}}(\beta)=\begin{cases} 
      0 & \|\beta\|_{1}\leq 1 \\
      \infty & \text{otherwise}
   \end{cases}.
\end{equation*}
We add a small ridge like $\ell_2$ penalty for Lasso and logistic Lasso (to ensure solutions). The randomized versions of these queries on data are described in details in \cite{harris2016selective}. We provide results for selection on instances of randomizations with two different distributions- the Gaussian and Laplace distribution. 
\medskip

\emph{\bf Data generating mechanism in simulations:} The entries $X_{ij}$, $1\leq i\leq n$, $1\leq j\leq p$, of predictor matrix $X$ are drawn independently from a standard normal and the columns of $X$ are normalized. The response vector $y\in\mathbb{R}^n$ is simulated from $\mathcal{N}(0,I_n)$ in the case of a Lasso and FS. For logistic loss $y_i\overset{i.i.d}{\sim}\textnormal{Bernoulli}(1/2)$, $1\leq i\leq n$. The true sparsity $s$ in all simulations is set to zero. Hence, we only need to check whether the constructed intervals cover zero while reporting coverages for the selected coefficients.

 In Lasso and logistic regression, the ridge penalty $\epsilon$ is set at $1/\sqrt{n}$ and the penalty level $\lambda$ is set to be the empirical average of $c\|X^TZ\|_{\infty}$, where $Z\sim\mathcal{N}(0,I_n)$ in the case of Lasso and $Z_i\overset{i.i.d.}{\sim}\textnormal{Bernoulli}(1/2)$, $1\leq i\leq n$, in the case of logistic regression. The computation of the tuning parameter is an empirical estimation of the theoretical value of $\lambda$ in \cite{negahban2009unified} that recovers the true underlying model. The size of selected set can vary with the value of constant $c$. 
We implement all three selection procedures, described in Table \ref{table:models} with randomization $\omega\sim\mathcal{N}(0, I_p)$ (Table \ref{table:gaussian:randomization}) and $\omega_i\overset{i.i.d.}{\sim}\textnormal{Laplace}(0,1)$, $1\leq i\leq n$, (Table \ref{table:laplace:randomization}) for dimensions $n=1000, p=500$. We compare the coverages of the selective intervals with naive confidence intervals using \eqref{eq:glm:target:asymptotics} that ignore selection bias; each reported as an average over $200$ iterations and with the target coverage $90\%$. We present the lengths of both selective intervals and naive ones; the lengths of the selective ones are comparable to the naive ones that highlight the high inferential power associated with the randomized procedures. All the code used here is available online.

\begin{table}[h!]
\centering
\caption{\textit{Gaussian randomization},  $n=1000, p=500$: }
\setlength{\tabcolsep}{7pt}
 \begin{tabular}{||c c c  c c c||} 
 \hline
  & \multicolumn{2}{c}{coverage} & \multicolumn{2}{c}{length} & \\
  & selective & naive  & selective  & naive  & $c$ \\ [0.5ex] 
 \hline\hline
 Lasso & 0.88 & 0.22 & 4.44 & 3.25 & $1.2$ \\ 
 \hline
 Logistic & 0.90 & 0.66 & 7.34 & 6.68 & $1.7$ \\
 \hline
 FS & 0.91 & 0.12 & 4.61 & 3.28 &  NA \\
 \hline
\end{tabular}
\label{table:gaussian:randomization}
\end{table}

\begin{table}[h!] 
\centering
\caption{\textit{Laplace randomization},  $n=1000, p=500$: }
\setlength{\tabcolsep}{7pt}
 \begin{tabular}{||c c c c c c||} 
 \hline
  & \multicolumn{2}{c}{coverage} & \multicolumn{2}{c}{length} & \\
  & selective & naive & selective & naive & $c$ \\ [0.5ex] 
 \hline \hline
 Lasso & 0.87 & 0.72 & 3.30 & 3.26 & $1.5$\\ 
 \hline
 Logistic & 0.89 & 0.85 & 6.57 & 6.62 & $3.2$ \\
 \hline
 FS & 0.91 & 0.77 &  2.94 &  3.27 & NA \\
 \hline
\end{tabular}
\label{table:laplace:randomization}
\end{table}

\medskip
\emph{\bf A sparse high dimensional example:} 
We draw predictor matrix $X\in \real^{n \times p}, n=500, p=5000$ with Gaussian entries (as described above) once. Simulate $Y\in \real^n$ as $Y\lvert X \sim X_E \beta_E + \epsilon, \; \epsilon \sim \mathcal{N}(0, I_n)$ in each draw of experiment where $|E|=5$. We use two signal regimes- low and moderate: (LS) with $5$ equally spread signals between $[0.5, 3.5]$ and (MS) with again $5$ signals spaced between $[3.5, 6.5]$. This is a deviation from the all noise generative mechanism. We compare coverages and lengths of intervals for the population coefficients post a randomized Lasso query based on the approximate pivot and the untruncated pivot in Table \ref{table:signal:regime}. 

\begin{table}[H]
\centering
\caption{\textit{Sparse model}, $n=500, p=5000$: }
\setlength{\tabcolsep}{7pt}
 \begin{tabular}{||c c c c c c||} 
 \hline
  & \multicolumn{2}{c}{coverage} & \multicolumn{2}{c}{length} & \\
  & selective & naive & selective & naive & $c$ \\ [0.5ex] 
 \hline \hline
(LS) & 0.88 & 0.33 & 4.47 & 3.34 & $1.1$\\ 
 \hline
 (MS) & 0.89 & 0.43 & 4.48 & 3.35 & $1.1$ \\
 \hline
\end{tabular}
\label{table:signal:regime}
\end{table}

\medskip
\emph{\bf HIV drug resistance analysis:} We conclude with inference for a protease inhibitor subset of the data analyzed in \cite{zhang2005comparison}, post solving a Gaussian-randomized Lasso using the theoretical value of tuning parameter with $c=1$ on the same. We select a model with active predictors of size $|E|=26$ from a set of $p=91$ potential mutations set for one of the drugs, Lamivudine (3TC). The sample consists of $n=633$ patients. We compute the selection adjusted confidence intervals based on our approach and compare it to the naive intervals based on normality of the least squares estimator for the selected coefficients. The below plot compares the adjusted and unadjusted inference, with the error bars representing the confidence intervals and the bar heights depicting the selective MLE and the unadjusted one, naive least squares estimator. 
\begin{figure}[h!] %
    \centering
    	\includegraphics[height= 8cm, width=13cm]{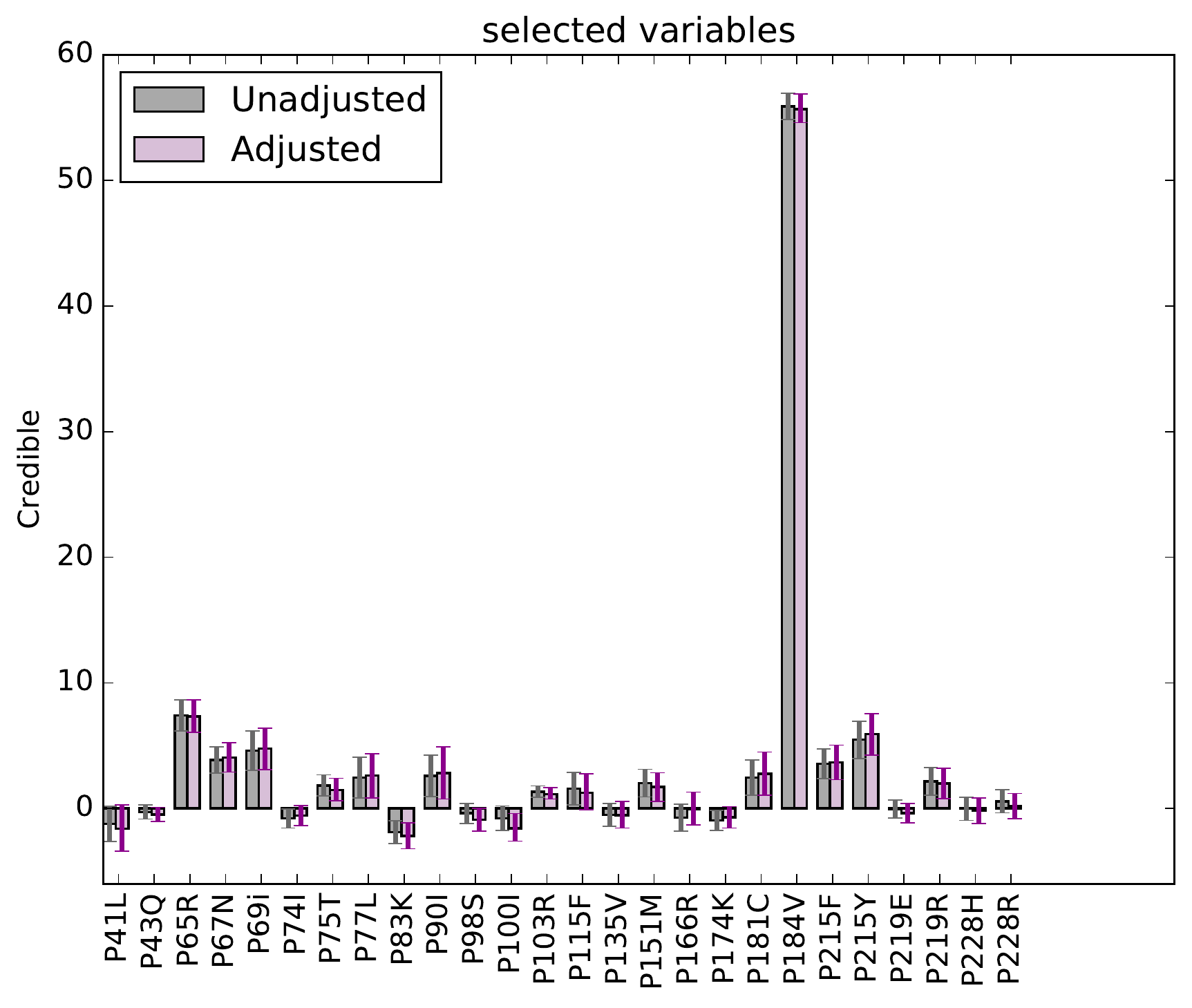}
    \caption{The grey bars depict unadjusted inference; pink bars depict adjusted inference based on the approximate selective law. With added randomization, the adjusted intervals are comparable in length to the unadjusted ones. Mutations \textbf{P174K, P115F, P74I} are no longer statistically significant, based on corrected inference using our approach. The selective MLE for the above mutations is shrunk more towards $0$ as compared to the unadjusted MLE; again an impact of overcoming the selective bias.}%
    \label{fig:HIV}%
\end{figure}
\begin{figure}[h!] %
    \centering
    	\includegraphics[height= 8cm, width=13cm]{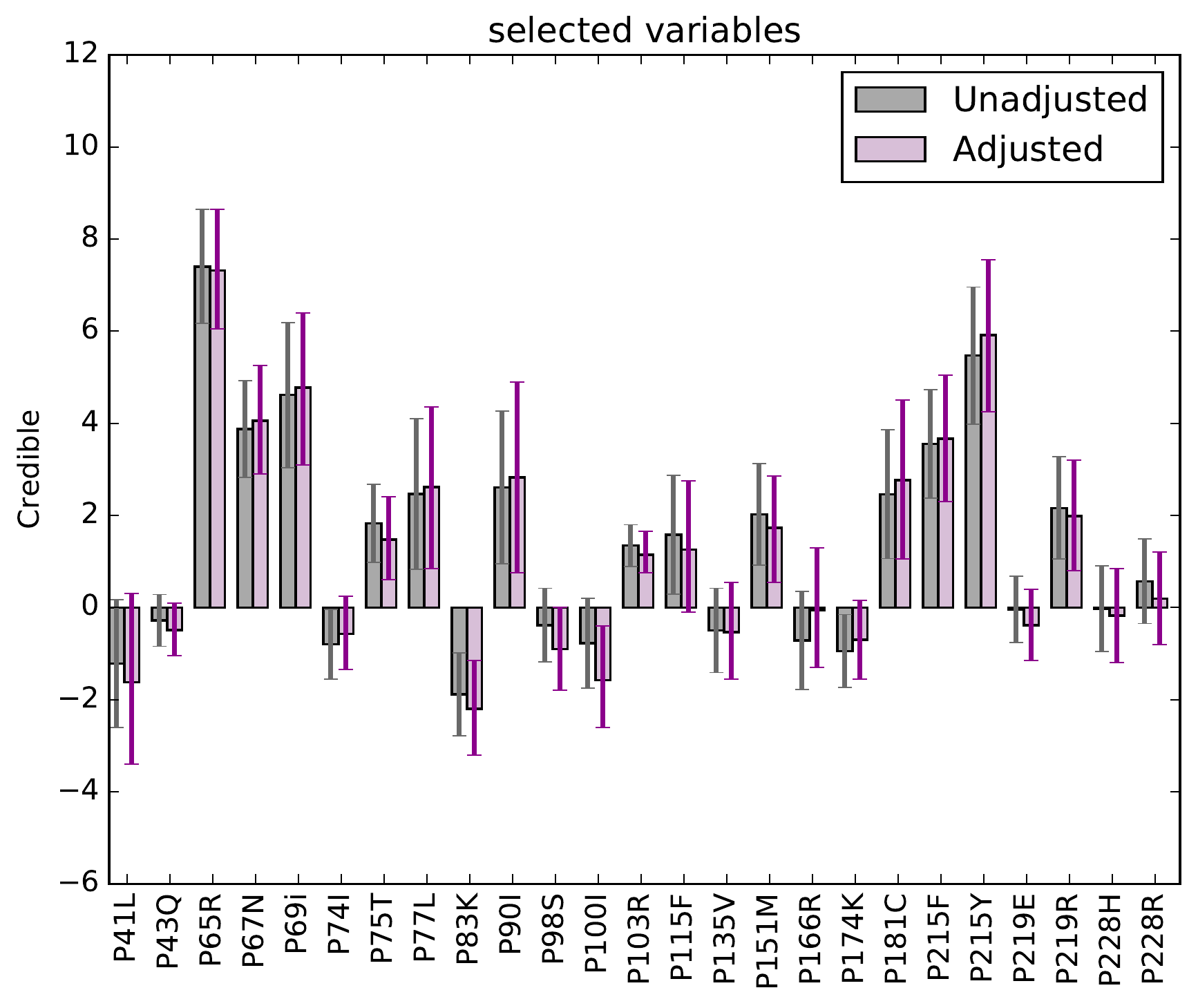}
    \caption{A more clear plot of Figure \ref{fig:HIV}, leaving out the largest effect size corresponding to mutation \textbf{P184V}.}%
    \label{fig:LG}%
\end{figure}

\section{Conclusion}
The contribution of this work is the proposal of an approximate pivot that can be used for valid inference post selection via a wide range of randomized algorithms. The main highlights of this approach are its ability to scale in high dimensions with the intervals possessing both valid coverage properties and higher statistical power. We see extensions of our approach as future directions to inference post selection of groups of variable by solving algorithms like the group lasso, explored in \cite{loftus2015selective, yang2016selective} and to bootstrapped versions of the approximate pivot. 


\clearpage
\bibliographystyle{plainnat}
\bibliography{references}


\newpage
\section{Appendix} \label{sec:appendix}

\subsection{KKT details} \label{sec:KKT}

Let us introduce some more notation before describing randomization reconstruction map from \eqref{canonical:randomized:program} in detail. 
Recall that $\bar{\beta}_E\in\mathbb{R}^{|E|}$ denote the solution of the unpenalized and non-randomized version of \eqref{canonical:randomized:program} including only the variables in $E$ ($X_E$ with rows $x_{i,E}$, $i=1, \ldots, n$), i.e.
\begin{equation*}
\begin{aligned}
	\bar{\beta}_E=\underset{\beta\in\mathbb{R}^{|E|}}{\textnormal{arg min}}-\sum_{i=1}^n(y_i\log\pi(x_{i,E}^T\beta)+(1-y_i)\log(1-\pi(x_{i,E}^T\beta))),
\end{aligned}
\end{equation*}
hence $\bar{\beta}_E$ satisfies $X_E^T(y-\pi(X_E\bar{\beta}_E))=0$.\footnote{A function $\pi$ is applied component-wise to a vector.}
Let us also write the following quantities
\begin{equation*}
\begin{aligned}
	\pi_E(\beta) &= \frac{\exp(X_E\beta)}{1+\exp(X_E\beta)} \\
	W_E(\beta) &= \textnormal{diag}\left(\pi_E(\beta)(1-\pi_E(\beta)\right) \\
	Q_E(\beta) &= X_E^T W_E(\beta) X_E \\
	c_{j,E}(\beta) &= X_{-E}^T W_E(\beta) X_E
\end{aligned}
\end{equation*}
for any $\beta\in\mathbb{R}^{|E|}$. 
The subgradient equation of \eqref{canonical:randomized:program} becomes
\begin{equation} \label{eq:glm:subgradient}
\begin{aligned}
	\nabla\ell\left(\begin{pmatrix}O_E \\ 0 \end{pmatrix};(X, y)\right)+  \begin{pmatrix} \lambda s_E \\ O_{-E}\end{pmatrix} -\omega+\epsilon\begin{pmatrix}
		O_E \\ 0
	\end{pmatrix} = 0,
\end{aligned}
\end{equation}
with the constraints $O\in\mathcal{K}$.
Taylor expansion of the gradient of the loss gives
\begin{equation*}
\begin{aligned}
	\nabla\ell\left(\begin{pmatrix} O_E \\ 0 \end{pmatrix}; ( X, y)\right)\approx -X^T(y-\pi_E(\bar{\beta}_E)) +  X^TW_E(\bar{\beta}_E) X_E(O_E-\bar{\beta}_E),
\end{aligned}
\end{equation*}
hence the subgradient equation \eqref{eq:glm:subgradient} becomes
\begin{equation*}
\begin{aligned}
	&\omega = \omega(D,O)= -\begin{pmatrix}
	Q_E(\bar{\beta}_E) &  0 \\ c_{j,E}(\bar{\beta}_E) &  I_{p-|E|}
	\end{pmatrix} D  + \begin{pmatrix} Q_E(\bar{\beta}_E)+\epsilon I \\ c_{j,E}(\bar{\beta}_E) \end{pmatrix}O_E + \begin{pmatrix} \lambda s_E \\ O_{-E} \end{pmatrix}
\end{aligned}
\end{equation*}
with the constraint $O\in \mathcal{K}$. Thus, we have
\begin{equation*}
\begin{aligned}
	A_0= -\begin{pmatrix}
	Q_E(\bar{\beta}_E) &  0 \\ c_{j,E}(\bar{\beta}_E) &  I_{p-|E|}
	\end{pmatrix}, \; B=\begin{pmatrix} Q_E(\bar{\beta}_E)+\epsilon I & 0 \\ c_{j,E}(\bar{\beta}_E) & I_{p-|E|} \end{pmatrix}, \;\; \gamma =\begin{pmatrix}
		\lambda s_E \\ 0
	\end{pmatrix}.
\end{aligned}
\end{equation*}

\subsection{Proofs} \label{sec:proofs}

\begin{proof-of-theorem}{\ref{thm:upper:bound}}

Denote $\phi_p(\cdot)$ as the density of a normal random variable $\mathcal{N}(0,I_p)$. Conditional on the data (fixing $D$ to be realized value of statistic), using a standard change of measure based on \eqref{randomization:reconstruction}, we have
\begin{equation*}
\setlength{\jot}{10pt}
\begin{aligned}
&\mathbb{P}(O \in \mathcal{K}\lvert T_{j\cdot E}= t_{j\cdot E}) \\
&= \int\limits_{\text{diag}(s_E){o_E}\geq 0} \phi_E\left(\frac{A_{j,E}t_{j\cdot E} + B_E o_E + c_{j,E}}{\tau}\right)\int\limits_{\|o_{-E}\|_{\infty}\leq 1}\phi_{p-E}\left(\frac{\alpha(o_E; t_{j\cdot E}) + o_{-E}}{\tau}\right)do_{-E}do_E \\
&=\int\limits_{\text{diag}(s_E){o_E}\geq 0} \phi_E\left(\frac{A_{j,E}t_{j\cdot E} + B_E o_E + c_{j,E}}{\tau}\right)\cdot \exp(H(o_E; t_{j\cdot E}))do_E \\
&=\mathbb{E}\left[\exp(H(O_E; t_{j\cdot E})) \mathbb{I}_{\{\text{diag}(s_E){O_E}\geq 0\}}\:\big|\: T_{j\cdot E}= t_{j\cdot E}\right] \\
&=\mathbb{E}\left[\exp(H(O_E; t_{j\cdot E})-\beta^T O_E) \exp(\beta^T O_E)\mathbb{I}_{\{\text{diag}(s_E){O_E}\geq 0\}}\:\big|\:T_{j\cdot E}= t_{j\cdot E}\right]\\ 
&\leq \mathbb{E}\left[\exp\left(\underset{\text{diag}(s_E){u}\geq 0}{\sup}\left\{H(u; t_{j\cdot E})-\beta^T u\right\}\right)\exp(\beta^T O_E)\:\Big|\: T_{j\cdot E}= t_{j\cdot E}\right]\\
&= \exp\left(\underset{\text{diag}(s_E){u}\geq 0}{\sup}\left\{H(u; t_{j\cdot E})-\beta^T u\right\}\right)\cdot\mathbb{E}\left[\exp(\beta^T O_E)\:\Big|\: T_{j\cdot E}= t_{j\cdot E}\right]
\end{aligned}
\end{equation*}
for all $\beta\in\mathbb{R}^{|E|}$.
Taking the logarithm of the selection probability and optimizing over $\beta\in\mathbb{R}^{|E|}$ gives us
\begin{equation*}
\begin{aligned}
\log\mathbb{P}(O\in\mathcal{K}|D) 
&\leq -\sup_{\beta}\inf_{\text{diag}(s_E){u}\geq 0}\left\{\beta^T u -H(u; t_{j\cdot E})- \log \mathbb{E}\left[\exp(\beta^T O_E)\:|\:D\right]\right\} 
\end{aligned}
\end{equation*}
which proves the claim in \ref{thm:upper:bound}.
\medskip

We apply an approximate minimax equality; approximate since an exact minimax result holds for a compact, convex selective region. In our case, the selective region takes the form 
\[\mathcal{K}_E =\{o_E: \text{diag}(s_E) o_E \geq 0\}\] which is convex, but not compact. We can however, work with a sufficiently large compact subset of $\mathcal{K}_E$  on which the active coefficients are supported with an almost measure $1$; thereby allowing us to relax the compactness assumption. Using minimax, we obtain the approximation for $\log\mathbb{P}(O \in \mathcal{K}\lvert D)$ as
\begin{equation*}
{\sup\limits_{{\text{diag}(s_E) o_E>0}} -\left\{\cfrac{\| A_{j,E}t_{j\cdot E} + B_E o_E + c_{j,E} \|_2^2}{2\tau^2} - H(o_E; t_{j\cdot E})\right\}.}
\end{equation*}

\end{proof-of-theorem}


\noindent \emph{\bf Details of smooth approximation in \eqref{barrier:approx}}:
Denoting $\mathcal{K}_E= \{o_E: \text{diag}(s_E) o_E>0\}$
\[  \chi_{\mathcal{K}_E}(o_E)=\begin{cases} 
      0 & \text{ if }  \text{diag}(s_E) o_E>0 \\
      \infty & \text {otherwise},
   \end{cases}
\]
we have
\begin{equation*}
\begin{aligned}
&\inf_{{\text{diag}(s_E) o_E>0}} \left\{\cfrac{\| t_{j\cdot E} A_{j,E}+ B_E o_E + c_{j,E} \|_2^2}{2\tau^2}- H(o_E; t_{j\cdot E})\right\}\\
&= \inf_{o_E} \left\{\cfrac{\| t_{j\cdot E} A_{j,E}+ B_E o_E + c_{j,E} \|_2^2}{2\tau^2}- H(o_E; t_{j\cdot E}) + \chi_{\mathcal{K}_E}(o_E)\right\}\\
&\approx \inf_{o_E} \left\{\cfrac{\| t_{j\cdot E} A_{j,E}+ B_E o_E + c_{j,E} \|_2^2}{2\tau^2}- H(o_E; t_{j\cdot E}) + \mathcal{B}(o_E)\right\}
\end{aligned}
\end{equation*}
where $\mathcal{B}(o_E)$ is a smooth version of discrete penalty $\chi_{\mathcal{K}_E}(o_E)$, with $\mathcal{B}(o_E) = \infty \text{ for } o_E \notin \mathcal{K}_E$ decaying to a $0$ penalty as continuously as we move deep into the selection region.


\begin{proof-of-lemma}{\ref{sel:MLE}}

The negative logarithm of the pseudo likelihood as a function of $b_{j\cdot E}$ in \eqref{pseudo:law}
\begin{equation*}
\begin{aligned}
-\log \hat{f}(t_{j\cdot E} & \lvert \hat{E} = E, F_{j\cdot E} ) = \frac{1}{2\sigma_j^2}(t_{j\cdot E} - b_{j\cdot E})^2+ \log{\sum_{t\in \mathbb{R}}\exp\left(-\frac{1}{2\sigma_j^2}(t- b_{j\cdot E})^2\right)\cdot \hat{h}(t)}\\
&= \frac{t_{j\cdot E}^2}{2\sigma_j^2} - \frac{t_{j\cdot E}b_{j\cdot E}}{\sigma_j^2} + \log{\sum_{t\in \mathbb{R}}\exp\left(-\frac{t^2}{2\sigma_j^2}+ \frac{t b_{j\cdot E}}{\sigma_j^2}\right)\cdot \hat{h}(t)}.
\end{aligned}
\end{equation*}
Setting the derivative of the above expression with respect to $b_{j\cdot E}$ to zero, we have that selective MLE satisfies \ref{sel:MLE}.
\end{proof-of-lemma}

\end{document}